\documentstyle[preprint,tighten,prc,aps]{revtex}
\date{\today}
\everymath={\displaystyle}
\begin{document}
\title{Scalar meson exchange and the baryon spectra}
\author{P. Stassart and Fl. Stancu}
\address{Universit\'{e} de Li\`ege, Institut de Physique, B 5, Sart 
Tilman,
B-4000 Li\`ege 1, Belgium}
\author{J.-M. Richard}
\address{Institut des Sciences Nucl\'eaires, Universit\'e Joseph
Fourier--IN2P3-CNRS,\\ 53, avenue des Martyrs, F-38026 Grenoble Cedex, 
France}
\preprint{\parbox{4cm}{ISN 99--36\\ nucl-th/9905015}}
\maketitle
\begin{abstract}
We explore the role of a scalar meson exchange interaction between 
quarks in a semirelativistic constituent quark model where the quarks 
are also subject to a linear confinement.  We search for a variational 
solution and study how the spectrum evolves when the strength of the 
scalar meson exchange increases.  Our results have good implications 
on the description of the low-lying baryon masses, especially on the 
long-standing problem of the relative position of the first 
positive-parity (Roper resonance) and the first orbitally excited 
states.
\end{abstract}

\vspace{1cm}

It has been suggested \cite{MA84,GL96a} that for energies below the 
scale of spontaneous chiral symmetry breaking, nonstrange and strange 
baryons can be viewed as systems of three quarks interacting via 
exchange of Goldstone bosons (pseudoscalar mesons).  It has also been 
shown that a properly parametrized interaction provides good baryon 
spectra with a correct order of positive and negative parity levels 
both in a nonrelativistic \cite{GL96b} or in a semirelativistic 
\cite{GL97} treatment.  In both parametrizations, the pseudoscalar 
meson exchange interaction has two distinct parts: a long-range Yukawa 
potential tail and a short-range part having opposite sign as compared 
to the Yukawa potential tail.  It is the latter which plays a major 
role in describing the baryon spectra in the frame of Goldstone boson 
exchange (GBE) models.

The underlying symmetry of the GBE model is related to the flavour--spin
$\mathrm{SU}_F(3) \times \mathrm{SU}_S(2)$ group combined with the S$_3$ 
symmetry required 
by a system of three identical quarks. A thorough independent analysis
of this model,  recently performed for $L=1$ baryons, has shown 
\cite{HG98}
that within the chiral picture one can obtain more satisfactory fits to 
the
observed spectrum than with one-gluon exchange (OGE)
models.

Although the models \cite{GL96a,GL96b,GL97} are thought to be a 
consequence of the spontaneous chiral symmetry breaking, the chiral 
partner of the pion, the $\sigma$-meson, is not considered explicitly.  
One can think of having mocked up its contribution in the parameters 
of the Hamiltonian \cite{GL96b,GL97}, e.g., in the regularisation 
parameter of the short-range spin--spin term.  The price which could 
have been paid is the large role played by the $\eta'$ meson exchange 
which comes into the interaction with a strength $g_0^2/(4\pi)$ about 
equal \cite{GL97} or larger \cite{GL96b} than the strength 
$g_8^2/(4\pi)$ associated with the pseudoscalar octet $\left(\pi, K, 
\eta\right)$.

Meson exchange potentials including $\sigma$-exchange have been considered
in the so-called ``hybrid" models where both meson and
gluon exchanges contribute to the quark--quark interaction.
However these studies were mainly devoted to the baryon--baryon
interaction \cite{VFF,KS99} and little conclusion has been explicitly
drawn about the role of the $\sigma$-exchange to the baryon spectrum.
Moreover these models are all nonrelativistic.

Here we study the explicit role of a $\sigma$-exchange interaction by 
considering the following semirelativistic Hamiltonian
\begin{equation}
\label{Hamil}	
H_0 = \sum\limits_{i}^{} \left(m_i^2 + p_i^2\right)^{1/2} + \frac{1}{2}
\sqrt{\sigma} \sum\limits_{i<j}^{} r_{ij} -
\frac{g_{\sigma}^2}{4\pi}
\sum\limits_{i<j}^{} \left[
\frac{\exp (- \mu_{\sigma} r_{ij})}{r_{ij} } - 
\frac{\exp (- \Lambda r_{ij})}{r_{ij} } 
                    \right],
\end{equation}
where $r_{ij}=|\vec{r}_{ij}|=|\vec{r}_i - \vec{r}_j|$. The 
second term is the confinement potential with a string tension 
\cite{CA83}
\begin{equation}
\sqrt{\sigma} = 1 \; \mathrm{GeV} \; \mathrm{fm}^{-1}.
\end{equation}
The third term is the $\sigma$-exchange interaction, the form of
which corresponds to the Pauli--Villars regularisation. In this term 
we fixed the $\sigma$-meson mass to
$\mu_{\sigma} = 600\;$MeV. The coupling constant $g_{\sigma}^2/(4\pi)$ 
and the regularization parameter $\Lambda$ are taken as variable 
parameters.
The interaction is attractive inasmuch as we view $\sigma$ as
a pair of correlated pions, as in the nucleon--nucleon interaction.
In practice this is achieved whenever $\mu_{\sigma} < \Lambda$.
 
From the pion--nucleon coupling constant $g_{\pi NN}^2/4\pi \simeq$ 
14, one obtains a pion--quark coupling constant $g_{\pi qq}^2/(4\pi)$ = 
0.67 which has been used in \cite{GL96b,GL97}.  Assuming a 
$\sigma$--nucleon coupling $g_{\sigma NN}^2/4\pi \simeq$ 8 \cite{MA87} 
one obtains, by scaling, $g_{\sigma qq}^2/4\pi \simeq$ 0.4.  We found 
that the values used in the literature \cite{VFF,KS99} are larger.
A recent evaluation \cite{RB99} of the $\sigma$--quark coupling constant
from the two-pion exchange interaction between constituent
quarks leads to $g_{\sigma qq}^2/4\pi \simeq 1$.  In 
the following we study the role of the $\sigma$-meson exchange for 
various ranges of $g_{\sigma qq}^2/(4\pi)$ as shown below.

We search for a variational solution of the Hamiltonian (\ref{Hamil}). 
Its 
general form is
\begin{equation}
\psi_{n}\left(\vec{r}_{12},\vec{r}_{23},\vec{r}_{31}\right) = 
\prod\limits_{i<j}^{}
f\left(r_{ij}\right)\phi_n\left(\vec{r}_{12},\vec{r}_{23},
\vec{r}_{31}\right).
\end{equation}
where $f$ is an orbital two-body correlation function  
and $\phi_{n}$ with $n\neq 0$ includes  orbital excitations \cite{SS85}. 
The function $f$ is parametrized as
\begin{equation}
f\left(r\right) = r^{\delta} \exp\left\{-W\left(r\right)\gamma r - 
\left[1 -
W\left(r\right)\right]\gamma' r^{1.5}\right\},\\
\end{equation}
with
\begin{equation}
W\left(r\right) = \frac{1 + \exp\left(-r_0/a\right)}{1 + 
\exp\left[\left(r -
r_0\right)/a\right]}.
\end{equation}
The quantities $\gamma$, $\gamma'$, $a$, $r_0$ and $\delta$ are 
variational parameters.

It is instructive to discuss the limit $\Lambda \rightarrow \infty$ 
first.  In this case the second term in the $\sigma$-exchange 
interaction vanishes.  Such a simplified form can be justified through 
arguments related to the differences appearing in the scalar and 
pseudoscalar propagators \cite{JA89} in a Nambu--Jona-Lasinio model. From  
there one can argue that the Fourier transform would lead to a 
Yukawa {\sl plus} a contact term for a scalar exchange, and a Yukawa 
{\sl minus} a contact term for a pseudoscalar exchange.  The latter is 
consistent with Refs.~\cite{GL96b,GL97}.  Due to the addition of the 
long and short range contributions in the scalar case, the two 
contributions can be combined in a single attractive term.  In the 
minimization procedure for $\Lambda \rightarrow \infty$, we found that 
the ground-state expectation value of (\ref{Hamil}) varies smoothly 
with all parameters but $\delta$.  This is quite natural because 
$\delta$ cares for the behaviour of the wave function near the origin, 
typical for the solution of a relativistic equation with a singular 
potential.  The dependence of the first three expectation values of 
(\ref{Hamil}) on the coupling constant is displayed in 
Fig.~\ref{Fig1}.  Alternatively, Table~\ref{tab:ener} shows the 
energies of the first excited states $E_{1p}$ and $E_{2s}$ for a 
ground state energy $E_{1s}$ set to 940 MeV. One can see that the energy 
difference between the first radially excited state and the negative 
parity state is positive up to the coupling constant value of 0.20 but 
tends to vanish.  Note also that the difference between the radially 
excited state and the ground state remains practically constant as a 
function of the coupling constant, while the mass difference between 
the orbitally excited state and the ground state increases with 
$g_{\sigma}^2/(4\pi)$.  The reason for this is that at 
$g_{\sigma}^2/4\pi \neq 0 $, every $s$ state is lowered with respect 
to the $p$ states.  The wave function of the latter is small around 
the origin, so that it reduces some of the attraction in the 
expectation value.

Our results show that the mass difference $E_{2s} - E_{1p}$
becomes  
negative for  $g_{\sigma}^2/4\pi \gtrsim 0.20$.
This is precisely the desired behaviour
for reproducing the correct order of the experimental spectrum, 
as in Ref.~\cite{GL96b}. These calculations indicate that a model
incorporating a potential, the Laplacian of which is negative in a  
region around the origin, can yield the right ordering
of the lowest positive and negative parity states. 
For the potential of Eq.~(\ref{Hamil}) such a situation is achieved
for suitable values of the coupling constant $g_{\sigma}^2/(4\pi)$.
Indeed the sign of the Laplacian governs the relative magnitude of 
radial vs.\ orbital excitations, as shown rigorously in the two-body 
case \cite{BMG} and approximately in the three-body case \cite{JMR}.

Realistically we expect $\Lambda$ to be finite.  We illustrate this 
case by taking $\Lambda = 2\;$GeV, consistent with the cut-off used in 
the Bonn nucleon--nucleon potential \cite{MA87}.  In this case, we let 
the coupling constant vary up to 1.34, the latter being 
twice the value of the pion coupling constant.  Fig.~\ref{Fig2} shows 
that the expectation value of 
the $L = 1$  state decreases slightly slower than that
of the first radially excited state.  
The latter remains practically parallel to the expectation value of the 
ground state.

It is important to understand the effects incorporated in the 
Hamiltonian (\ref{Hamil}). For the sake of the discussion, we introduce 
the quantity
\begin{equation}
	\label{def:R}
R = ( E_{1p} - E_{1s} )/( E_{2s} - E_{1s} )
\end{equation}
where $E_{1s}$, $E_{1p}$ and $E_{2s}$ are the expectation values for 
the ground state, the first negative parity state and the first 
radially excited state, respectively.  So that if $R \gtrsim 1$, the 
desired order is obtained.  It is well known that the ``naive" 
potentials of harmonic oscillator type have $R < 1$ (see 
e.g.\cite{IK}), and no tuning could improve the situation.  It is also 
known that the addition of the OGE interaction with reasonable 
strength does not improve the situation either, the Roper resonance 
always appearing above the first negative parity states, in 
contradiction to the experimental situation.  The fact that $R < 1$ 
for ``naive" potentials is not surprising, at least for the 2-body 
problem.  There are general theorems stating that $R$ is smaller than 
unity if $\triangle V \geq 0$ \cite{BMG}.  This happens for most 
empirical potentials and in particular for the celebrated Coulomb + 
linear potential of heavy quarkonia.  Although there do not exist 
rigorous theorems in the 3-body case, within the approximation of the 
lowest hyperspherical partial wave, one can also show that $R < 1$ if 
$\triangle V \geq 0$.  \cite{JMR}.

Actually we studied in detail several distinct effects contributing to 
the position and ordering of levels:

{\sl i)} In our potential we introduce components with $\triangle V 
\leq 0$ through the Yukawa-type potential.  This is the case of 
Fig.~\ref{Fig1} or Table~\ref{tab:ener}.  When a regularizing term 
with a finite $\Lambda$ is subtracted from the Yukawa type potential 
it is not surprising that $R$ decreases because the regularization 
term leads to a $\triangle V \geq 0$ contribution at small values of 
$r$.

{\sl ii)} We adopt a relativistic kinematics.  In the 2-body case 
(Herbst equation) we find that $R$ is closer to unity with a 
relativistic kinematics than if we treat the same problem 
nonrelativistically (NR).  For clarity in Table~\ref{tab:R} we show 
results obtained in the 2-body case with the interaction
\begin{equation}
V = \lambda r - g \exp{(-\mu r)}/r,
\end{equation}
where the parameters $\lambda$, $g$ and $\mu$ are taken in units of 
the quark mass m.  One can see that with or without Yukawa-type 
potential, the ratio $R$ is systematically larger in the 
semirelativistic case than in the nonrelativistic case. The case 
$\lambda=0.01$ tests that relativistic effects disappear in the 
weak-coupling limit.

{\sl iii)} We looked comparatively at the 2- and 3-body cases to see 
if there are differences in the value of $R$.  In the NR case it was 
shown\cite{JMR} that $R$ is very similar for the 2-body and the 3-body 
problem for potentials of type $ V \propto r^\beta$ with $\beta \geq 
-1$.  It occurs however that $R$ might be larger in the 3-body case 
than in the 2-body one when the potential $V$ is such that $\triangle 
V \leq 0$.  In our case, this is more delicate.  We have a linear term 
with a positive Laplacian, and a Yukawa term with a negative Laplacian.  
The latter being however of short-range character, it is 
less easily felt by baryons which, in first a approximation, obey a 
two-body dynamics with an effective angular momentum $J=3/2$ 
\cite{JMR}.  For instance, for the non-relativistic version of our 
Hamiltonian (\ref{Hamil}) with $m_{i}=1$, $\sqrt{\sigma}=2$, 
$g^{2}_{\sigma}/(4\pi)=0.2$, $\mu_{\sigma}=1$ and $\Lambda=\infty$, 
the ratio $R$ is found equal to $R\simeq0.67$ for  the 3-body case, to 
be compared to $R\simeq0.60$
the 2-body case.

An interesting finding is that the three effects above, $\triangle V 
\leq 0$, the relativistic kinematics and the 3-body effect cumulate 
together and lead to the result shown in Tables~\ref{tab:ener} 
and~\ref{tab:R}.

A correct level ordering of the first positive parity and the first 
orbitally excited states has been achieved in Refs.~\cite{GL96b,GL97} 
not at the level of the spin-independent Hamiltonian --- as discussed 
above --- but through adding the GBE interaction.  As mentioned in the 
beginning of this study, this interaction contains flavour--spin 
operators, i.e., it is related to the $\mathrm{SU}_F(3) \times 
\mathrm{SU}_S(2)$ group.  In a simpler version, one can use 
$\mathrm{SU}_I(2)$ instead of $\mathrm{SU}_F(3)$.  Then one can easily 
understand the role of the GBE interaction by considering the 
following operator
\begin{equation}
\label{def:oper}	
	O^{IS} = - \sum\limits_{ i\ <\ j}^{} {\vec{\tau}}_{i} \cdot 
{\vec{\tau}}_{j} \
{\vec{\sigma}}_{i} \cdot {\vec{\sigma}}_{j},
\end{equation}
for a three-quark system.  The effect of this interaction is similar 
to that shown in Fig.~\ref{Fig1} as suggested by Table~\ref{tab:exp} 
where three expectation values of $O^{IS}$ are indicated.  One can see 
that the ground state and the first radially excited state, both of 
orbital symmetry $[3]_O$ are lowered by 15 units while the first 
negative parity state of either spin 1/2 (i.e., $[21]_S$) or 3/2 (i.e., 
$[3]_S$) are shifted up or down by a quantity five times smaller.  Then 
the ground state and the Roper resonance are both lowered while the 
negative parity states are much less affected by the GBE interaction, 
leading in practice \cite{GL96b,GL97} to a good level ordering.

In Ref.~\cite{RB99} it was found that the two-pion exchange plays also 
a significant role in the quark-quark interaction. The obtained 
interaction
has a spin independent central component, which, averaged over the
isospin part of the nucleon wave function, gives rise to an attractive
spin independent interaction. The $\sigma$-exchange represents in fact
a good approximation of such a central interaction. Here we have shown 
that
a scalar attractive interaction cooperates together with  
the flavour-spin operator (8). In this respect our findings are  
consistent with those of Ref.~\cite{RB99}.\par 

Other important results of Ref.~\cite{RB99}, relevant for hadron
spectroscopy, are that the spin-orbit components of the two-pion exchange
interaction provide a cancelling mechanism for the spin-orbit
interaction resulting from the confinement potential and that its tensor
component cancels out the one-pion exchange tensor component.\par 

The conclusion is that Goldstone boson exchange models should include 
explicitly the chiral partner of the pseudoscalar mesons.  It might be 
misleading to mock up its effect in some nontrivial parameters of the 
model.  Our next step will be to study the case of a potential with 
both scalar and pseudoscalar terms.  However the question still 
remains about the role of the chromomagnetic interaction.


\acknowledgements 
We are grateful to L.~Glozman and W.~Plessas for very useful discussions.
We also thank C.~Semay and F.~Brau for providing some eigenvalues of the
Herbst equation obtained with their method \cite{BS98} for testing our 
variational algorithm, and M.~Mangin-Brinet and J.~Carbonell for 
checking accurately some 3-body calculations and for interesting 
discussions.


%

\listoftables
\listoffigures
\begin{table}
	\caption{\label{tab:ener}Energies (in MeV) of the first orbitally 
excited state $E_{1p}$
and the first radially excited state $E_{2s}$ 
obtained variationally for a $3q$ system described by the Hamiltonian 
(\ref{Hamil})
with $\Lambda \rightarrow \infty$. The last column gives the values of
$R$ defined by Eq.~(\ref{def:R}).}
\begin{center}
\begin{tabular}{cccc}
$g_{\sigma}^2/(4\pi)$\,\, &  $E_{1p}$ & $E_{2s}$ & $R$ \\
\hline
\, 0.00 &  \,\,\, 1287  &   \,\,\, 1444 & 0.688 \\
\, 0.05 &  \,\,\, 1309  &   \,\,\, 1444 & 0.732 \\
\, 0.10 &  \,\,\, 1333  &   \,\,\, 1444 & 0.780 \\
\, 0.15 &  \,\,\, 1380  &   \,\,\, 1445 & 0.871 \\
\, 0.20 &  \,\,\, 1431  &   \,\,\, 1446 & 0.970 \\
\end{tabular}
\end{center}
\end{table}

\begin{table}

\caption{\label{tab:R}Values of $R$, Eq.~(\ref{def:R}), for the 
nonrelativistic (NR) and the semirelativistic (SR) 2-body equation 
with the potential (7).  The parameters $\lambda$, $g$ and $\mu$ are 
dimensionless, as explained in the text}
\begin{tabular}{dcccc}
$\lambda$ & g & $\mu$ & NR & SR \\
\hline 
0.01 & 0 & - & 0.585 & 0.588 \\
0.865 & 0 & - & 0.585 & 0.628 \\
2.0   & 0 & - & 0.585 & 0.644 \\
0.865 & 0.2 & 1.765 & 0.597 & 0.655 
\end{tabular}
\end{table} 

\begin{table}
    \caption{\label{tab:exp}The expectation value of the interaction 
    (\ref{def:oper}) for a colour singlet 3q system}
\begin{center}
%
\caption{\label{Fig1}
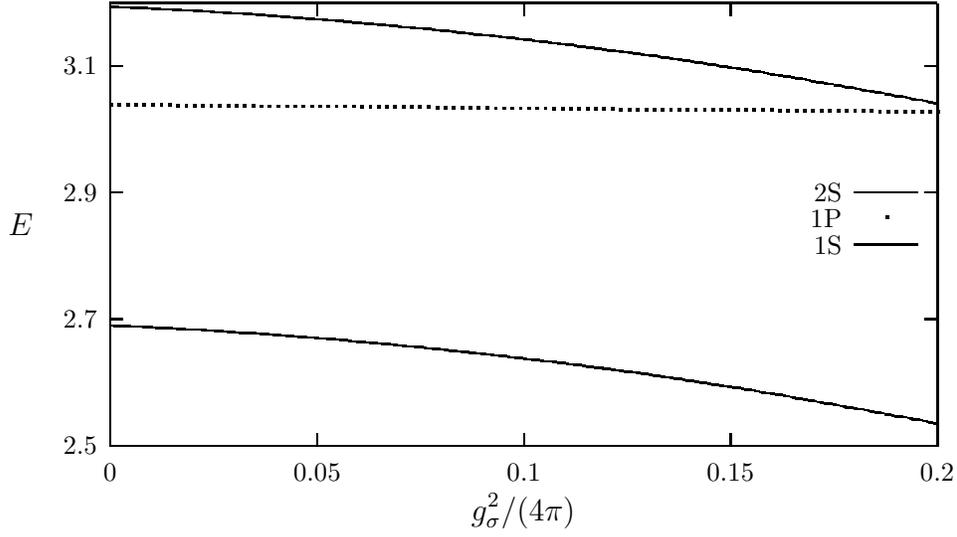
Energies of the first three eigenvalues of Hamiltonian (\ref{Hamil}) 
for increasing values of the coupling constant $g_{\sigma}^{2}/(4\pi)$ 
in the limit $\Lambda \rightarrow \infty$.}
    \end{figure}

\begin{figure}
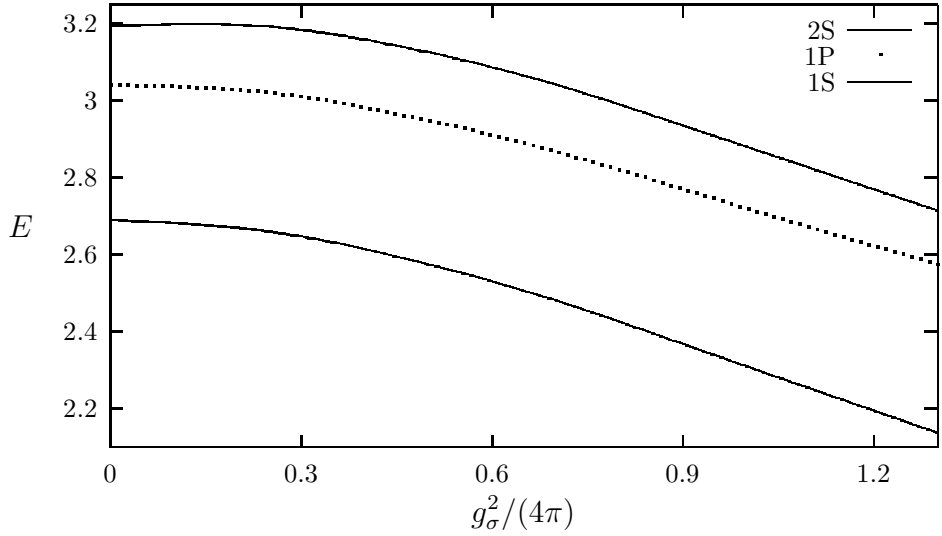

\setlength{\unitlength}{0.240900pt}
\ifx\plotpoint\undefined\newsavebox{\plotpoint}\fi
%
    
\caption{\label{Fig2}
Same as Fig.~\ref{Fig1} but for a finite cut-off $\Lambda = 2\;$GeV. }
\end{figure}
\end{document}